\documentclass[preprint2]{aastex}
\begin{document}

\lefthead{ }
\righthead{ }

\def \vjec{\vfill\eject}
\def\hmpc{MpcH^{-1}}
\def\chisq{$\chi^2$}
\def \m3{{\rm Mark III}}
\def \etal {{\it et al.\ }}
\def \cf {{\it cf.\ }}
\def \vs {{\it vs.\ }}
\def \via {{\it via\ }}
\def \ie {{\it i.e.\ } }
\def \eg{{\it e.g.\ }}
\def\kms{km s^{-1}}
\def\br{{\bf r}}
\def\bv{{\bf v}}
\def\bV{{\bf V}}
 
\def\plotfour#1#2#3#4{\centering \leavevmode
\epsfxsize=.24\columnwidth \epsfbox{#1} \hfil
\epsfxsize=.24\columnwidth \epsfbox{#2} \hfil
\epsfxsize=.24\columnwidth \epsfbox{#3} \hfil
\epsfxsize=.24\columnwidth \epsfbox{#4}}

\newcommand{\pa}{\partial}
\newcommand{\cri}{_{\rm cr}}
\newcommand{\zi}{ \mbox{\boldmath$\xi$} }
\newcommand{\Bl}{ \left( }
\newcommand{\Br}{ \right)}
\newcommand{\Blsq}{ \left[ }
\newcommand{\Brsq}{ \right] }
\newcommand{\CH}{ V_H^{2} }
\newcommand{\CB}{ GM_B }
\newcommand{\RBB} {\Bl \frac{R_B}{R_H} \Br} 
\newcommand{\CBB} { \frac{GM_B}{R_H V_H^{2}}  }

\title{GALAXY FORMATION IN TRIAXIAL HALOES:\\ BLACK HOLE-BULGE-DARK HALO 
    CORRELATION} 
\author{Amr A.  El-Zant}
\affil{Center for Astrophysics \& Space Astronomy, Campus Box 391, University
   of Colorado, Campus Box 391, Boulder, CO 80309-0391, and\\ Department of
   Physics \& Astronomy, University of Kentucky, Lexington, KY 40506-0055,
   USA \\ email: {\tt elzant@pa.uky.edu}}

\author{Isaac Shlosman\altaffilmark{1} and Mitchell C. Begelman}
\affil{Joint Institute for Laboratory Astrophysics, University of Colorado,
   Campus Box 440, Boulder, CO 80309-0440, USA\\ email: {\tt
   shlosman@pa.uky.edu, mitch@jila.colorado.edu}} 

\altaffiltext{1}{JILA Visiting Fellow. Permanent address: Department of
Physics \& Astronomy, University of Kentucky, Lexington, KY 40506-0055, USA}

\and
\author{Juhan Frank}
\affil{Department of Physics \& Astronomy, Louisiana State University,
   Baton Rouge, LA 70803-4001, USA\\ email: {\tt
   frank@rouge.phys.lsu.edu}}   

\begin{abstract}
The masses of supermassive black holes (SBHs) show correlations with bulge
properties in disk and elliptical galaxies. We study the formation of galactic
structure within flat-core {\it triaxial} haloes and show that these correlations 
can be understood within the framework of a baryonic component modifying the
orbital structure in the underlying potential. In particular, we find that
terminal properties of bulges and their central SBHs are constrained by the
destruction of box orbits in the harmonic cores of dark haloes and the emergence 
of progressively less eccentric loop orbits there. SBH masses,
$M_\bullet$, should exhibit a tighter correlation with bulge velocity dispersions, 
$\sigma_{\rm B}$, than with bulge masses, $M_{\rm B}$, in accord with observations, 
if there is a significant scatter in the $M_{\rm H}-\sigma_{\rm H}$ relation for 
the halo. In the context of this model the observed $M_\bullet-\sigma_{\rm B}$ 
relation implies that haloes should exhibit a  Faber-Jackson type relationship 
between their masses and velocity dispersions. The most important prediction of 
our model is that halo properties determine the bulge and SBH parameters. The 
model also has important implications for galactic morphology and the process
of disk formation.
\end{abstract}


\newpage

\section{Introduction}

The possibility that supermassive black holes (SBHs) inhabit the centers of 
many if not most galaxies, and the observed correlation between SBH
masses and galactic bulge properties,\footnote{The term ``bulge'' in this 
paper will refer to bulges of disk galaxies, or to elliptical galaxies, when
no disk is  present.} has potentially a fundamental significance for our
understanding of galaxy formation and evolution. The relationships between
black hole and bulge properties include a loose relationship between SBH
and bulge masses, $M_\bullet \sim 0.001 M_{\rm B}$, and an apparently much 
tighter one between the SBH mass and the velocity dispersion in the 
corresponding bulge, $M_\bullet \sim \sigma_{\rm B}^4$ (e.g., Ferrarese \& 
Merritt 2000; Gebhardt et al. 2000; Tremaine et al. 2002; cf. reviews by 
Kormendy \& Gebhardt 2001; Merritt \& Ferrarese 2001). 

In this paper we  attempt to provide a physical explanation for these relationships 
between SBHs and their host galaxies. Our model is based on the interaction between 
the dark haloes of
galaxies and the baryonic components settling in their midst. As baryonic matter 
accumulates to form the bulge and SBH, the orbital structure of the underlying 
gravitational potential is modified. This, in turn, affects the subsequent 
accumulation of gas, which is highly dissipative and therefore is sensitive to 
orbital geometry. Consequently, halo properties will determine those of the bulge 
and the SBH, and this will give rise to correlations between them that are
compatible with the observed ones.

Cuspy, triaxial haloes appear to be a natural outcome of dissipationless
collapse in a  cold dark matter (CDM) dominated universe (e.g., Cole \& Lacey
1996). Interactions with the baryonic component during the initial stages of
collapse  may affect the  triaxiality, making it milder but still non-negligible
(Dubinsky 1994).  Before it even becomes dominant in the central regions,  a
clumpy baryonic component can also level off  the central cusps of  dark
haloes, producing harmonic cores (El-Zant, Shlosman \& Hoffman 2001). Even
though this may affect the equidensity contours,  again making them rounder,
it need not symmetrize the  equipotentials; these can remain asymmetric
if triaxiality is not affected beyond some radius (for example, a homogeneous
bar has a non-axisymmetric force contribution  inside its density figure).
From an observational standpoint it appears that haloes of
fully formed galaxies tend to have nearly constant density cores 
(de Blok \& Bosma 2002) and that residual potential axial ratios of about 0.9 
in CDM haloes are plausible, {\em even in present day galaxies} (e.g., Kuijken 
\& Tremaine 1994; Rix \& Zaritsky 1995; Rix 1995).

The orbital structure of the inner regions of slowly rotating, non-axisymmetric  
potentials with harmonic cores is dominated by box orbits (e.g., Binney \&
Tremaine 1987), which have no particular sense of circulation.  They are 
self-intersecting and, therefore, cannot be populated with gas. Dissipation
causes material to sink quickly toward the only long-lived attractor 
available  --- the center (e.g., Pfenniger \& Norman 1990; El-Zant 1999).  
In the process, interaction
with the triaxial harmonic core causes the baryonic material to lose most
of its angular momentum. The combined system would, therefore, also be slowly
rotating. Thus, unless star formation terminates the collapse, the final
concentration of the first baryonic material could be extremely
large.

The onset of star formation is expected to occur when the baryonic
material becomes self-gravitating --- roughly speaking, when its density
becomes larger than that of the halo core.  This also happens to be the
criterion for the destruction of the harmonic core and the emergence of loop
(or tube) orbits, which do have a definite sense of circulation. The 
role of the SBH is to contribute to the emergence of these orbits 
in the very central region. It is the collusion between the SBH and 
the more extended hot baryonic component in creating the loop orbits
that leads to the correlations claimed in this paper.

Whereas, in three dimensions, a box orbit can be represented as a superposition 
of three (generally) incommensurable  radial oscillations along mutually perpendicular
axes, which are efficiently attenuated by dissipation, loop orbits are best 
described in terms of modest radial and vertical excursions superposed on 
rotational motion about the center. While the vertical and radial excursions, 
like the oscillations characterizing box orbits, are attenuated by dissipation, 
the rotational motion is not efficiently dissipated among gas clouds populating 
such orbits (in the same sense of circulation). 

The minimization of the radial and vertical oscillations results in closed periodic 
orbits that are confined to a plane.
We expect the amount of gas dissipation on loop orbits to depend on their axial
ratios --- for, again, motion on highly eccentric orbits would lead to shocks and the
accompanying loss of angular momentum on a short dynamical timescale,
as observed in numerical simulations of gas flows in barred galaxies (e.g., Heller 
\& Shlosman 1994). Lacking a detailed model for the dissipation rate associated with gaseous 
motion, we will assume that there is a critical eccentricity {\it below} which 
the loop orbits can serve as long-lived attractors for dissipative motion.
In other words, gas populating these orbits will evolve only secularly
and not dynamically. Fortunately, this dependence on critical eccentricity will turn
out to be weak.

In the following we explore, within  the above framework, the formation of galactic 
bulges and central compact objects. We show that it is possible to deduce a well-defined 
linear correlation between the masses of the central SBHs and corresponding bulges in given 
dark matter haloes, with mass ratios comparable to those observed.  We demonstrate that, 
if dark matter halo cores follow a Faber-Jackson type relation (Faber \& Jackson 1976) 
between their masses and velocity dispersions, then a similar relation also applies 
to the bulges. An  $M_\bullet-\sigma_{\rm B}$ relation between the SBH mass and the bulge 
velocity dispersion, which under certain assumptions can have smaller scatter  
than the $M_\bullet-M_{\rm B}$ relation, also follows naturally.  Within this framework, 
it is possible to make a number of testable predictions concerning the related  
structures of bulges, SBHs and
their host haloes. We discuss these in the final section. Formal aspects of the 
perturbation analysis applied to bulge-halo systems have been deferred to Appendices~A and B.
Preliminary results of this work have been reported by Shlosman (2002).

\section{The model} 

Since, at this stage, we are interested in {\em generic} dynamical
phenomena related to orbital structure, the exact form of the halo
potential is immaterial --- as long as it exhibits a harmonic core where no
loop orbits can exist. The exact distribution  of the baryonic material is
also not crucial, except, again, for its central density distribution. If this
diverges, say as $\rho \propto r^{-1}$, loop orbits will be created all the way to 
the center, since the bulge-halo system no longer possesses a harmonic core. In this
case there is no need to form the SBH. If, on the other hand,
the (proto)-bulge has a harmonic core of its own, there will still be a 
nearly constant density region near the center with only box orbits  --- unless a 
central point mass is present. We will assume that `cuspy' bulges, when they exist, 
are products of later evolution (for example, a result of star formation and 
cold dissipationless collapse: e.g., Lokas \& Hoffman 2000).  

We use a logarithmic standard form (e.g., Binney \& Tremaine 1987) to represent 
the halo potential:
\begin{equation}
\Phi_{\rm H} = \frac{1}{2} V^{2}_{\rm H} \log (R_{\rm H}^{2} +x^{2} + \beta^{-2}
    y^{2} + \gamma^{-2} z^2),  
\label{logarithm}
\end{equation}
where  $V_{\rm H}$ is the asymptotic (in the limit $R \gg R_{\rm H}$ and $\beta, \gamma 
\rightarrow 1$) circular velocity, $R_{\rm H}$ is the core 
radius  and  $\beta$, $\gamma  < 1$ are the potential axis ratios.
We will consider the process of bulge formation
to be terminated, or at least substantially slowed down, when the combined 
(baryonic plus halo) potential admits sufficiently round non-intersecting loop orbits, 
which permit the long-term circulation of gas without excessive  dissipation. 
Such motions necessarily take place in a symmetry plane determined by the 
(orbit-averaged) angular momentum
(e.g., Frank, King \& Raine 2002).  
Therefore, for our purposes, it will suffice to consider 
only orbits in the plane $z = 0$ and to ignore the 
vertical dimension. We fix $\beta$ at $0.9$ (though the effects of its variation are
discussed where relevant) and adopt a threshold axial 
ratio $p_{\rm crit}$ for orbits that can be populated with gas. 
The value of $p_{\rm crit}$ that best describes when this happens remains to be investigated, 
but its mere existence is what is important here. As we show below, our results are 
rather insensitive to the exact value of $p_{\rm crit}$.  

The halo core mass  will be taken to be $M_{\rm H} = V_{\rm H}^2 R_{\rm H}/G$ and 
we define the halo density to be $\rho_{\rm H} = M_{\rm H}/R_{\rm H}^3 = 
V_{\rm H}^2 / R_{\rm H}^2  G$. This is 
conveniently close to the value of the density in the region where the 
potential is effectively harmonic --- that is, within the region where, in the absence
of the bulge component, no loop orbits  exist for $\beta = 0.9$. 
We note, however, that a bulge with a larger core radius probes a region of the halo 
core with smaller average density than the region probed by a bulge with a relatively 
small core radius. This effect will be discussed in Section~4.

Since our aforementioned criterion depends on the potential in a chosen
symmetry plane, the exact three-dimensional mass distribution of the bulge is unimportant,
as the same planar potential may arise from a variety of these.  A particularly convenient 
form for the potential in this situation is that of Miyamoto \& Nagai (1975),
\begin{equation}
\Phi_{\rm D} = -  \frac{GM_{\rm D}} {\sqrt{ x^{2}+y^{2}+ \Bl A +\sqrt{B^{2}+z^{2}} \Br^{2}} }.     
\label{miyamoto}
\end{equation}
This potential can approximate a disk-bulge system, with 
the parameters $A$ and $B$ determining the scale length and height,
respectively. In the symmetry plane, therefore,  the potential 
can arise from a range of density distributions --- from 
highly flattened to spherical ones. In general, when we refer to the
``bulge'' we will have a spherical system in mind with $A=0$. In particular,
all densities quoted are calculated under this assumption. We will
comment on the effects of assuming a flattened distribution where 
relevant. In that case $M_{\rm D}$ in eq.~(\ref{miyamoto}) 
is replaced by $M_{\rm B}$, the spherical bulge mass. The bulge {\em core}
is defined by the radius $R_{\rm B}=A+B$ and encloses the core mass
$M_{\rm BC} = 2^{-3/2} M_{\rm B}$. The bulge core density is
defined as $\rho_{\rm B} = M_{\rm BC}/R_{\rm B}^{3}$.

The scaling relations that arise do not depend on the absolute masses and lengths, but 
instead, on the ratios of these quantities among the various contributions
to the potential.  An important property of the dynamics in our model is that
{\em any transformation that leaves the relative masses and lengths
of the components constant will also not affect the orbital structure once
scale transformations are taken into account.}  Such transformations only
change the time units, hence the dynamics remains invariant on a new characteristic 
timescale. In addition, {\em in regions where the baryonic and 
halo components have the same mass distributions, i.e., inside their cores,
only the density ratios  of these components determine  their relative 
contributions to the potential,} and hence the orbital structure.

For the remainder of this paper, we will describe all components of the potential 
in terms of units scaled to the dimensions of the halo potential, eq.~(\ref{logarithm}). 
We scale all distances to $R_{\rm H}$, with $\hat R_{\rm B} \equiv R_{\rm B}/ R_{\rm H}$, 
and all masses 
to $M_{\rm H}$ so that $\hat M_{\rm B} \equiv M_{\rm B}/ M_{\rm H}$.  The normalized 
bulge core density is then $\hat \rho_{\rm B} \equiv 2^{-3/2} \hat M_{\rm B}/ 
\hat R_{\rm B}^3 $. The bulge potential in scaled units is 
\begin{equation}
\Phi_{\rm B} = -  V_{\rm H}^2 \frac{2^{3/2} \hat\rho_{\rm B} \hat R_{\rm B}^3 } 
    {\sqrt{x^{2}+y^{2}+ \hat R_{\rm B}^2} }.     
\label{bulgepot}
\end{equation}
The black hole potential, in the same units, is 
\begin{equation}
\Phi_{\bullet} = -  V_{\rm H}^2 \frac{2^{3/2} K \hat\rho_{\rm B} \hat R_{\rm B}^3 } 
    {\sqrt{ x^{2}+y^{2}} },         
\label{BHpot}
\end{equation}
where $K \equiv M_\bullet / M_{\rm B}$. Note that all orbits in the total potential 
$\Phi = \Phi_{\rm H} + \Phi_{\rm B} + \Phi_\bullet$ depend only on three parameters, 
which may be taken to be  $\hat R_{\rm B}$, $\hat \rho_{\rm B}$, and $K$.

\section{Bulge formation in triaxial haloes and the  $M_\bullet-M_{\rm B}$ correlation}

In this section, we demonstrate that the arguments outlined above imply a linear 
relation between the SBH and bulge masses, provided that the baryonic and CDM density 
distributions are not completely axisymmetric and exhibit, at some stage, nearly 
constant density cores.  

\subsection{Critical eccentricities and the emergence of SBH-bulge correlations}

\subsubsection{The inner minimum and the role of bulge density}

\begin{figure}[ht!!!!!!!!!!!!!!!!!!!!!!!!!]
\vbox to2.5in{\rule{0pt}{2.5in}}
\includegraphics{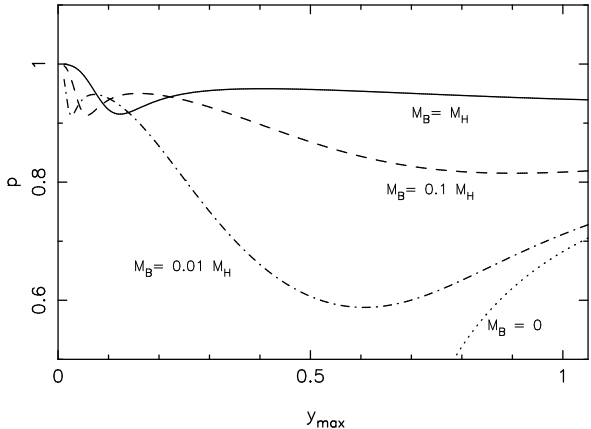}
\caption{Axis ratios of closed loop orbits $p=x_{\rm max}/y_{\rm max}$ as a function of 
$y_{\rm max}$. The dotted line corresponds to a system with only a halo contribution 
(i.e., no bulge and no SBH), the solid line refers to a system with a ``bulge'' in 
the form of a Plummer sphere with $\hat M_{\rm B} = 1$ and $\hat R_{\rm B} = 0.4$, 
the dashed line represents a bulge with $\hat M_{\rm B} = 0.1$ and $\hat R_{\rm B} 
= 0.19$, while the dashed-dotted line represents a bulge with $\hat M_{\rm B} = 0.01$ 
and a radius of $\hat R_{\rm B} = 0.086$. In all cases with the bulge present, 
$\hat \rho_{\rm B} = 5.52$ and an SBH with a fraction $K=10^{-3}$ of the 
bulge mass is present.  
}
\label{homofin}
\end{figure}

Fig.~\ref{homofin} (the result of orbital integrations) exhibits the axis 
ratios of the closed loop orbits $p= x_{\rm max}/y_{\rm max}$ as a function of the 
longer axis length ($y_{\rm max}$). In this figure, we have fixed the scaled bulge 
core density, $\hat\rho_{\rm B} = 5.52$, and the SBH-to-bulge mass ratio, 
$K= M_\bullet/M_{\rm B} = 10^{-3}$, but vary the scaled bulge core radius, $\hat R_{\rm B}$.  
Consequently, the bulge masses $\hat M_{\rm B}$ range over a factor of $100$.  As expected, 
when only the halo contributes to the potential (dotted line), $p$ quickly drops to 
zero inside the halo core, since no loop orbits can exist there. (For halo 
potential axis ratios smaller than the adopted value of 0.9, i.e., greater triaxiality, 
$p$ reaches 0 at radii that are progressively closer to the halo core radius, 
$y_{\rm max} \rightarrow 1$. As the baryonic
component, both  in the bulge and the SBH, becomes progressively more massive, 
correspondingly rounder loop orbits appear. Loop orbits in the very central
region are produced by the presence of the SBH. The latter is represented numerically 
by a Plummer sphere with a softening scale of $0.002$, in scaled units. Near the 
center, where the SBH contribution is dominant, the axis ratios of the loop orbits 
tend to $p \rightarrow 1$.

The axis ratio curves exhibit two distinct minima.  It is apparent that the inner 
minima of the curves in Fig.~\ref{homofin} have (nearly) the same values of $p$, 
but occur at different radii $\propto R_{\rm B}$. This is a consequence of assuming a 
fixed ratio of SBH-to-bulge mass, $K$, and a constant bulge core density 
$\hat \rho_{\rm B} > 1 $.  Because the bulge and halo densities are nearly uniform in 
the inner regions, their relative contributions to the potential are determined by 
their density ratio. 

In the absence of the SBH contribution, $p$ in Fig.~\ref{homofin} would tend to zero 
within the effective harmonic core of the bulge-halo system (see eq.~[\ref{expo}]). 
When the relative density is kept constant (and $>1$) this is proportional to 
$\hat R_{\rm B} \sim \hat M_{\rm B}^{1/3}$ (the effect of density variation will 
be examined in Section~\ref{inmin}). 
The reason for this is that inside the nearly constant-density bulge core, the 
potential, which is now a superposition of two nearly harmonic potentials, is 
nearly harmonic (even if the new core is less triaxial than the halo alone, e.g., 
if the bulge is assumed to be spherical).  The inner minima of the curves in 
Fig.~\ref{homofin}, therefore, correspond to a transition from the region where the 
SBH provides the dominant contribution toward the creation of loop orbits to that 
where the bulge provides this contribution. Therefore, the minima occur at radii 
where the gravitational acceleration due to the SBH is proportional to that due 
to the bulge+halo, $R_\bullet \sim (M_\bullet/M_{\rm B})^{1/3} R_{\rm B} = 
K^{1/3} R_{\rm B}$ (since the minimum is located well within the bulge core radius). 
Since $K$ is taken to be constant, we expect the minima to occur at radii
$\propto R_{\rm B}$. 

In Section~2, we have defined the critical value $p_{\rm crit}$ above which gas 
circulation can be sustained for secular timescales.  
Here we have shown that given a critical value, $p_{\rm crit}$, for the inner minimum
and the relative bulge-to-halo core density ratio, a value of $K$ associated
with this minimum is determined. However, satisfying the
condition $p > p_{\rm crit}$ at the inner minimum does not imply that
the same condition is satisfied at all radii within $R_{\rm H}$. Until
it is satisfied at the outer minimum, as well, further
evolution can occur. This is briefly sketched below and outlined in more detail 
in Section~3.4.

\subsubsection{Outer minimum, minimal bulge mass, and final black hole mass}

Until the critical value of $p$ is reached at all radii inside the halo core, 
the bulge will continue to grow, since outside the bulge core $p$ declines 
again. It is clear from Fig.~1 that, in contrast to that of the inner minimum,
the value of $p$ at the outer minimum depends more sensitively
on the bulge mass 
than on its density. The growth of the bulge, therefore, determines a {\em minimal} 
bulge mass fixed by the condition that $p \geq p_{\rm crit}$ at all radii 
within the halo core.

If the SBH did not continue to grow along with the bulge, this growth 
in the bulge mass would decrease the linear correlation coefficient $K$, but,
as we show in Section~3.4, this leads only to near-independence of the final
value of $K$ on $p_{\rm crit}$ and still keeps the values of $K$ within the
observed range. 
 
\subsection{Scaling relationships determined by inner minima}
\label{inmin}

As shown in the previous section, the creation of loop orbits with axis ratio above 
a given value depends, in the central region, solely on the ratio of the bulge-to-halo 
core density and the mass ratio $K = M_\bullet/M_{\rm B}$. The actual value of the
bulge mass determines only the position of the minimum (as a fraction of the halo 
core radius), not the axis ratio at the minimum. But even the radius of the minimum 
is largely insensitive to the bulge mass, i.e., $\propto M_{\rm B}^{1/3}$, provided that 
the minimum actually exists and the bulge core density relative to the halo harmonic 
core density remains constant. We will now examine the bounds on $M_\bullet$ and $K$ 
which ensure the existence of a  minimum $p \geq p_{\rm crit}$ inside the bulge 
core and observe the effect of varying the bulge density.

\begin{figure}[ht!!!!!!!!!!!!!!!!!!!!!!!!!]
\vbox to2.5in{\rule{0pt}{2.5 in}}
\includegraphics{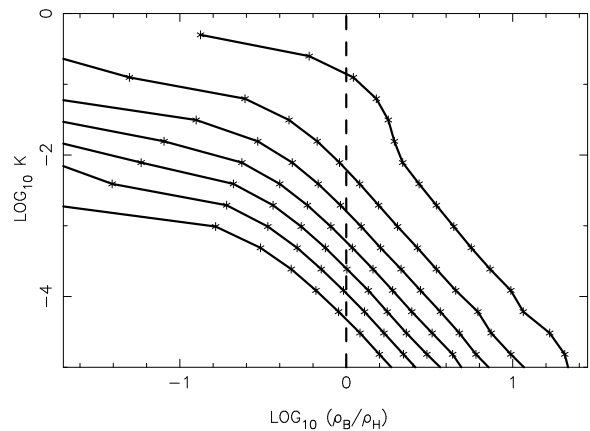}
\caption{SBH-to-bulge mass ratio $K$ for different $p_{\rm crit}$ as a function of the
bulge-to-halo core density ratio required to create closed loop orbits with axis ratios
$p \ge p_{\rm crit}$ {\it within} the bulge core, ${\hat R} < {\hat R}_{\rm B}$.
$M_{\rm B}$ is chosen to satisfy $p > p_{\rm crit}$ everywhere outside 
the bulge.  From left to right, the characteristic $p_{\rm crit}$ associated with 
a given curve increases from 0.3 to 0.9, in increments of 0.1.  The bulge core 
density ${\hat\rho}_{\rm B}$ is varied by fixing $\hat M_{\rm B}$ and changing 
${\hat R}_{\rm B}$. The vertical dashed line at $\hat{\rho}_{\rm B} = 1$ is the 
approximate boundary between the self- and non-self-gravitating regimes in the bulge. 
For densities ${\hat \rho}_{\rm B} \sim 1$, the $p_{\rm crit}=0.9$ curve corresponds 
to an (unrealistically) overmassive SBH and is shown
for comparison only. The range of $K$ values, for these densities, thus is limited to within
$10^{-4}-10^{-2}$.
}
\label{compac}
\end{figure}

We calculate and plot $K = K({\hat\rho}_{\rm B})$ for different choices of $p_{\rm crit}$ 
(Fig.~\ref{compac}), in the following way.  First, without any loss of generality, 
we choose a bulge mass $\hat M_{\rm B}$ that 
is large enough that the outer minimum satisfies $p \ge p_{\rm crit}$, for all 
$p_{\rm crit} \leq 0.9$. This is done in order to focus
on the effects of the inner minima only.
(Recall, from the previous section, that the value of the 
outer minimum depends on the mass of the bulge.)  
Fig.~\ref{compac}, therefore, exhibits the effect of ${\hat\rho}_{\rm B}$ and $K$ on the 
inner minima of the $p$ curves at ${\hat R} < {\hat R}_{\rm B}$. In order to obtain 
$K = K({\hat\rho}_{\rm B})$, we vary $M_\bullet$, which is some fraction $K$ of the chosen 
bulge mass. The density is varied by contracting the bulge (by decreasing its core 
radius) until the condition $p \ge p_{\rm crit}$ is satisfied everywhere inside the 
bulge core.  

One observes in Fig.~2 that  at smaller densities $K$ tends toward an asymptotic (and 
maximal) value associated with a given $p_{\rm crit}$. We refer to this hereafter as 
the ``asymptotic'' regime.  In the transition to this regime, along a $p_{\rm crit} = 
{\rm const.}$ curve, the value of $K$ is monotonically increasing. In other words, 
the bulge contribution toward the creation of loop orbits of a given eccentricity 
gradually diminishes, and is compensated by a greater contribution from the SBH 
component. In the process, the minimum in $p$ moves outward. The asymptotic regime 
corresponds to a situation in which the existence of loop orbits with the required 
elongation depends on the value of $K$, irrespective of the bulge density (that is, 
effectively, only on the SBH mass).  In this limit the SBH  contribution to the 
potential is sufficient to create loop orbits with $p \ge p_{\rm crit}$ at all radii 
within the halo core  (i.e., at ${\hat R} \leq 1$), without additional contributions 
from the bulge component (cf. Fig.~3). Essentially, it corresponds to a rapid collapse 
to the center and formation of the SBH by a large fraction of the baryonic material, 
bypassing the  formation of a bulge. This regime appears to be of academic interest 
only, as it implies ${\hat\rho}_{\rm B} < 1$ all the way to the center; it 
will not be discussed further.

\begin{figure}[ht!!!!!!!!!!!!!!!!!!!!!!!!!]

\vbox to2.5in{\rule{0pt}{2.5in}}

\includegraphics{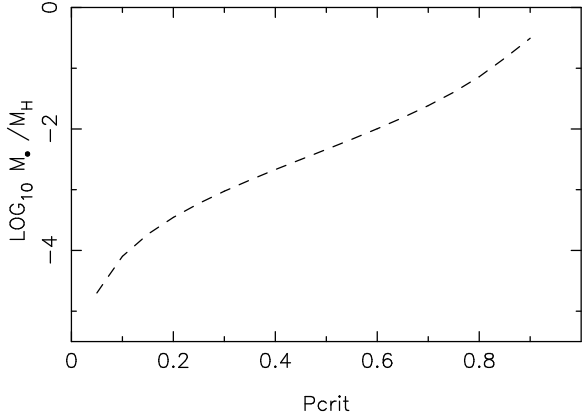}
\caption{${\hat M_\bullet}$ sufficient to create loop orbits rounder than a given 
$p_{\rm crit}$ without collapse of the baryonic material to form a bulge.}
\label{blacratloops}
\end{figure}   

A second regime characterizes the dynamical state of the SBH-bulge-dark halo system 
with ${\hat\rho}_{\rm B} \ga 1$.  In this ``scaling regime'' the 
$\log K - \log{\hat\rho}_{\rm B}$ curves  are parallel straight lines with average 
slopes of about  $-2.5$.  This can be explained in the following manner.
From eq.~(\ref{expo}) we know that the radius of the effective harmonic core of the 
bulge-halo system, $y_{\rm max}$, is proportional to $\hat R_{\rm B}^{5/2} 
\hat M_{\rm B}^{-1/2}$.  
In the scaling regime, the location of the minimum in $p$ will be proportional to 
$y_{\rm max}$.  Now, in order to maintain the minimum at a specified value of 
$p_{\rm crit}$, the gravitational acceleration due to the SBH must be proportional 
to the acceleration due to the halo (which contains the nonaxisymmetry) at $y_{\rm max}$.  
This implies $\hat \rho_{\rm B} \hat R_{\rm B}^3 K / y_{\rm max}^2 \propto y_{\rm max}$.  
Substituting for $y_{\rm max}$ and noting that $\hat \rho_{\rm B} \propto \hat M_{\rm B}/ 
\hat R_{\rm B}^3$, 
we obtain $K \propto \hat \rho_{\rm B}^{-5/2}$. This is approximately what is found 
from Figure~2 and it also holds if we change the mass of the bulge keeping the 
radius constant and again using eq.~(\ref{expo}).
Note, furthermore, that for constant $\hat \rho_{\rm B}$ this relation predicts the radius 
of the effective harmonic core (and the location of the minimum) to be proportional 
to $R_{\rm B}$, as expected from the heuristic considerations of the previous section.

While the asymptotic values of $K$ depend on $M_{\rm B}$, the values of $K$ in scaling
regime do not. The transition between these two regimes can, therefore, be
characterized by a sharp change in the behavior of $K$, as seen from Fig.~2.
 
We are mainly interested in the scaling region, because the onset of star formation 
can be tied to the baryonic component becoming self-gravitating, which corresponds 
roughly to the bulge core density exceeding that of the background halo, i.e., 
${\hat\rho}_{\rm B} \ga 1$. In this case, and for $p_{\rm crit}\la 0.8$, the SBH 
contributes significantly to the potential only at radii ${\hat R}_\bullet \ll 1$. 
The $p$ curves, therefore, exhibit a definite inner minimum well inside the halo core.

The values of ${\hat \rho}_{\rm B}$ and the initial $K$, in principle, also depend on the 
critical eccentricity, $p_{\rm crit}$, of the inner minimum, which is  expected to be 
independent of the bulge mass, but which does depend on complex gas dynamics. If, after 
the SBH forms, the bulge mass falls short of the value required
for  creating sufficently round closed loop orbits at all radii inside the halo core, 
it will continue to grow
until the outer minimum of the axis ratio curve also attains $p \ge p_{\rm crit}$. In
Section~3.4, we show that the SBH growth, if continued, becomes intermittent. In 
the next section, we also demonstrate that one can place constraints on the possible 
range of $K$ values by considering this constraint on the outer minimum. 

\subsection{Minimal bulge mass and bulge growth}

To obtain loop orbits rounder than a given $p_{\rm crit}$ {\em at all radii} ${\hat R} 
< 1$, and not only in the central region, one in fact needs to take into account 
the bulge-to-halo core mass ratio and not only the density ratio.  This can already 
be seen in Fig.~\ref{homofin}, where all curves with SBH and bulge contributions
exhibit an inner minimum with $p \ga 0.9$, while $p$ declines significantly as one 
moves further out. For low bulge masses, an additional outer minimum emerges, within 
${\hat R}_{\rm B} < {\hat R} < 1$, before $p$ rises again outside the halo harmonic core. 
Only one of the $p$ curves, corresponding to the most massive bulge, has $p > 0.9$ 
at {\em all} radii. Therefore, for a given halo, there is a minimal bulge mass, 
${\hat M}_{\rm Bmin}$, that is necessary to create sufficiently round loop orbits at all 
radii ${\hat R} > {\hat R}_{\rm B}.$

\begin{figure}[ht!!!!!!!!!!!!!!!!!!!!!!!!!]
\vbox to2.5in{\rule{0pt}{2.5in}}
\includegraphics{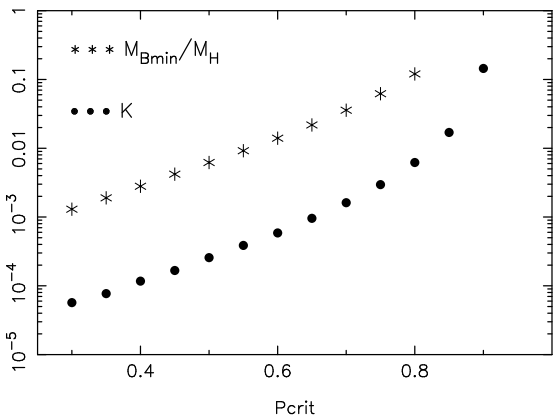}
\caption{Asterisks: bulge masses ${\hat M}_{\rm Bmin}(p_{\rm crit})$ required to 
produce loop orbits with $p \ge p_{\rm crit}$ at radii {\it outside} the bulge core 
${\hat R}_{\rm B}$, as a function of $p_{\rm crit}$ for ${\hat\rho}_{\rm B} = 1$. 
Filled circles: corresponding ratio of the SBH-to-bulge mass $K$ required to produce {\em 
inner minima} in the loop orbit axis ratios with $p \ge p_{\rm crit}$ as a function 
of $p_{\rm crit}$ at ${\hat\rho}_{\rm B} = 1$ (from Fig.~2). The $p_{\rm crit} > 
0.8$ points are shown for comparison only, as this condition requires 
(for $\hat \rho_{\rm B} \sim 1$) an unrealistically massive SBH, with 
strong effects on the potential at all radii within $R_{\rm H}$.  This 
accounts for the steep increase in $K$ for these values of $p_{\rm crit}$.
}
\label{complots}
\end{figure}

If we assume that the bulge core density varies at most by a factor of a few   
(${\hat\rho}_{\rm B}= O[1]$), ${\hat R}_{\rm B}$, which is only weakly dependent on
the bulge mass and density ($\hat R_{\rm B} \sim \hat M_{\rm B}^{1/3} 
{\hat \rho}_{\rm B}^{-1/3}$), varies little. The minimal 
bulge mass is also
nearly constant (calculations show, for example, that it varies by $\sim 30$\% when 
${\hat\rho}_{\rm B}$ changes from 1 to 2.5). 
There is more sensitivity to the assumed $p_{\rm crit}$. In Fig.~\ref{complots} 
(asterisks) we show the bulge mass required to create loop orbits with 
$p \ge p_{\rm crit}$ at all radii outside ${\hat R}_{\rm B}$, for 
${\hat \rho}_{\rm B} =1$.  
These are the masses necessary, at this density, to produce outer minima with the 
required values. For bulge densities ${\hat\rho}_{\rm B} \sim 1$ these minima lie at 
radii ${\hat R} \ga 0.5$. Thus, unless the SBH can contribute significantly to the 
potential at radii comparable to $R_{\rm H}$, which seems implausible, the dominant 
contributions to the potential at the radii examined here should be only those due 
to the bulge and halo components. Therefore, once the halo parameters are fixed, 
the creation of loop orbits with given $p \ge p_{\rm crit}$ outside the bulge core will 
depend only on the bulge parameters ${\hat R}_{\rm B}$ and ${\hat M}_{\rm B}$.

\begin{figure}[ht!!!!!!!!!!!!!!!!!!!!!!!!!]
\vbox to2.5in{\rule{0pt}{2.5in}}
\includegraphics{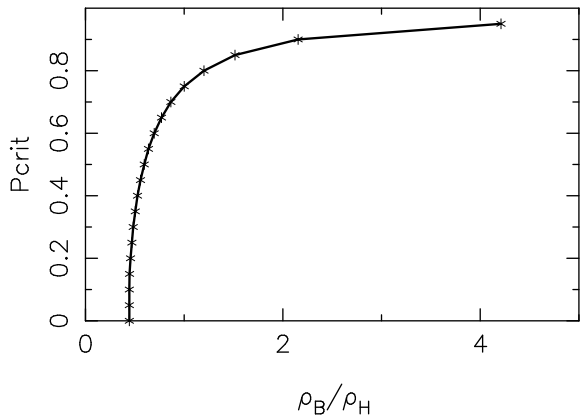}
\caption{Minimum density required for creating loop orbits with 
$p \ge p_{\rm crit}$ without appeal to an overmassive SBH (see text).
}
\label{exmags}
\end{figure}

For ${\hat \rho}_{\rm B} =1$, a well-defined inner minimum in the $p$ curve exists 
inside the bulge core only if $p_{\rm crit} \le 0.8$. For larger 
$p_{\rm crit}$, the inner minimum moves outward and merges with the outer one. 
Systems with these properties lie outside the scaling region in Fig.~2. In this 
case, for the condition $p \ge p_{\rm crit}$ to be satisfied, the SBH 
has to contribute significantly at all radii --- no matter how massive the bulge is.  
(For ${\hat \rho}_{\rm B} =1$, the curve with $p_{\rm crit}=0.9$ in Fig.~2 lies  
in this regime.)
Furthermore, the minimal bulge density required to create loop orbits with 
$p \ge p_{\rm crit}$, without appeal to an overmassive SBH,
increases rapidly when $p_{\rm crit} \ga 0.8$
(Fig.~\ref{exmags}), suggesting that the bulk of the material may form stars before 
the condition $p \ge p_{\rm crit}$ is reached everywhere within the halo core. 
If we demand that $\hat \rho_{\rm B} \leq 1$, then 
$p_{\rm crit }$ must be smaller than 0.8 in order for the model to be 
plausible, implying that $K$ must lie in the range $10^{-4} - 10^{-2}$.

On the other hand, star formation can be somewhat delayed to higher densities.
A reasonable range lies within ${\hat\rho}_{\rm B} \approx 1-10$. Still higher densities
in the bulge can be excluded based on the observed rotation curves. In this range of 
${\hat\rho}_{\rm B}$, values of $p_{\rm crit} > 0.8$ become feasible without invoking 
unrealistically massive SBHs, as seen from Fig.~\ref{exmags}. One may also
reasonably assume that, due to enhanced  dissipation, higher densities require
larger $p_{\rm crit}$ to maintain long-lived gaseous motion (detailed modeling, 
however, will be necessary to determine exactly how $p_{\rm crit}$
depends on the system parameters). In this situation, the range in $K$ that can be 
inferred from Fig.~2, is again about $10^{-4} - 10^{-2}$ and compatible with the observed
value of $\sim 10^{-3}$, which exhibits a significant scatter. 

The values of $K$ obtained so far, which should be regarded as {\it initial} values, are 
fixed by the inner minimum of the axis ratio curves.  As argued in Section 3.1, further 
evolution of the bulge will occur unless sufficiently round loop orbits also exist at 
the outer minimum.  In the next section we examine how this evolution can modify the 
range of $K$, if at all.

\subsection{Black hole growth and SBH-to-bulge mass ratio}

There are basic differences between the conditions for satisfying $p\geq p_{\rm crit}$  
at the inner and outer minima in Fig.~\ref{homofin}. As discussed in Section~3.1, 
the initial growth of 
the SBH is terminated after the inner loop orbits reach the critical eccentricity. The 
growth of the bulge, on the other hand, continues until the required $p_{\rm crit}$ is 
reached at all radii inside the halo core. 
Since the increase of bulge mass, at roughly constant density, moves the inner minimum 
outward as described in Section~3.1, the SBH would have to continue growing in 
pace with the 
bulge mass if it is to maintain the critical value of $K$ (which is determined 
only by the values of $\hat \rho_{\rm B}$ and $p_{\rm crit}$).  
Does this actually happen, and if so, how?

\begin{figure}[ht!!!!!!!!!!!!!!!!!!!!!!!!!]
\vbox to2.5in{\rule{0pt}{2.5in}}
\includegraphics{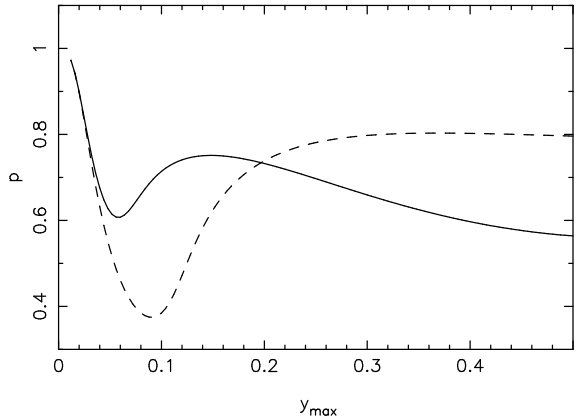}
\caption{Axis ratios of closed loop orbits $p=x_{\rm max}/y_{\rm max}$ as a function of
$y_{\rm max}$ (as in Fig.~1 but for ${\hat \rho}_{\rm B}=1$ and growing the bulge
mass without that of the SBH mass, thus decreasing $K$). 
The solid line corresponds to $K=10^{-3}$ and ${\hat M}_{\rm B} = 10^{-2}$. The dashed 
line corresponds to $K=10^{-4}$ and ${\hat M}_{\rm B} = 10^{-1}$. Note the `gap' which 
opens with decreasing $K$ and allows the gas to flow in again as $p$ becomes less
than $p_{\rm crit}$ in this region.
}
\label{homofin2}
\end{figure}

To see how the parallel growth of the SBH and bulge could come about, consider the 
solid curve in Fig.~\ref{homofin2}, which represents the state of the SBH/bulge system 
at the end of the initial stage of infall. This is equivalent to one of the curves (say, 
the dash-dotted line) in Fig.~1, except that we have chosen $p_{\rm crit} \approx 0.6$ for
the inner minimum, instead of 0.9. This 
line has two maxima: a left-hand one at the origin, determined by the SBH, and a 
right-hand one, determined by the bulge core. Note that the outer part of the solid curve 
drops below $p_{\rm crit}$, implying that the bulge will continue to grow. Now suppose 
this growth occurs {\it without} corresponding growth of the SBH (in contrast to the 
case in Fig.~1, where $K$ is kept constant). If the bulge continues to grow while the 
SBH growth is stopped, the right-hand maximum moves further to the right, while the 
left-hand maximum stays the same. This opens a widening `gap' at the position of the 
inner minimum which drops below $p_{\rm crit}$, allowing gas to flow again toward the 
radius of influence of the SBH, $R_\bullet$. This is illustrated by the dashed curve 
in Fig.~\ref{homofin2}, which corresponds to a factor of 10 increase in $\hat M_{\rm B}$ 
accompanied by a similar {\it drop} in $K$. The gas is expected to accumulate in this 
vicinity. However, once a substantial amount of gas has collected at $R_\bullet$, it 
becomes self-gravitating and is prone to global self-gravitating instabilities. The 
fastest of these instabilities, $m=2$ modes or bar instability (e.g., Bardeen 1975), 
have been discussed in a similar context by Shlosman, Frank \&  Begelman (1989). They 
induce rapid (dynamical) gas inflow. Hence we expect that the growth of the SBH at this 
stage will be intermittent, but because the time-averaged conditions for infall at the 
inner and outer maximum do not change, we  expect the value of $K$ to stay roughly 
constant within the range given in Fig.~2, namely $\sim 10^{-4}-10^{-2}$, depending on 
the value of $p_{\rm crit}$. 

Is there always enough gas to accumulate at $\sim R_\bullet$ in order to trigger a bar 
instability, due to the opening of the gap between the SBH and the bulge? One can imagine 
the opposite extreme to that discussed above, in which insufficient gas enters the 
widening gap from outside, perhaps because star formation is efficient during the early 
stages of infall. Let us suppose, for the sake of argument, that the SBH does {\em not} 
grow beyond its initial mass, as determined by the initial value of $K$. How does this 
affect the range of final $K$ values?

Figure~4 shows minimal bulge masses, ${\hat M}_{\rm Bmin}\sim 10^{-1}-10^{-3}$, for 
a range of $p_{\rm crit}$ values. While we have no estimate for the initial bulge 
masses which define the initial $K$ in Fig.~2, it is possible to rule out
very small masses, e.g., $< 10^{-3}$, because the total baryonic mass within the
halo core is taken to be about 10\%. This means that for lower $p_{\rm crit}\approx
0.3-0.4$, the initial bulge mass is equal to or larger than ${\hat M}_{\rm Bmin}$
estimated in Section~3.3,
and, therefore, will not grow beyond its initial value. It is easy to understand
this result, because one needs small baryonic masses to create loop orbits
with such large eccentricities. This means that in this regime the initial value
of $K\sim 10^{-4}$ (in Figures~2 and 4) is also its final value.

Alternatively, in the regime of larger $p_{\rm crit}\approx 0.7-0.8$, the bulge can
grow at most by a factor of $\sim 30-100$, reducing the initial $K$ by this amount.
Luckily, the initial values of $K$ for high $p_{\rm crit}$ lie around their high
end, $\sim 10^{-2}$. Reduction by up to two orders of magnitude brings them
again to about $10^{-4}$. Hence, whether the SBH grows after the initial stage or
not does not destroy the $M_\bullet-M_{\rm B}$ correlation and does not move the
values of $K$ outside the observed range. A corollary of this discussion is that 
the final value of $K$ appears insensitive to the value of $p_{\rm crit}$, if the
growth of the SBH is supressed, and depends strongly on $p_{\rm crit}$, if the
SBH grows in tandem with the bulge.  
Note also that the variation of ${\hat M}_{\rm Bmin}(p_{\rm crit})$ 
(represented by the asterisks in Fig.~4), follows closely that of  $K(p_{\rm crit})$ 
(represented by filled circles). The ratio of these two quantities, 
$M_\bullet M_{\rm H} / M_{\rm Bmin}^2 \sim 0.03$ is, therefore, largely independent of 
$p_{\rm crit}$. In other words, the minimal bulge mass is predicted to be about 5 
times the geometric mean between the SBH and the halo masses, provided that the SBH 
grows along with the bulge.

To summarize, the initial value of the black hole-to-bulge mass ratio
$K$ is fixed by the density at which the gaseous material
in the inner region becomes self-gravitating and forms stars, and by the value of 
$p_{\rm crit}$. However, unless loop orbits of sufficient  eccentricity are created
in the {\em whole} harmonic halo core region, the mass of the bulge will continue
to grow. At the same time, the dynamical infall onto the SBH can choke, 
if the remaining gas mass is insufficient to cause bar instability and channel it 
to the SBH. In the latter case $K$ will decline below its initial value as the bulge 
grows, but will tend to approach a final value roughly independent of $p_{\rm crit}$.  
In the opposite limit, the SBH will grow in proportion to the bulge, and $K$ will stay 
constant (as shown by Figs.~2 and 4). In either limit, we assume that the process stops 
when the mass of  the baryonic component reaches the minimal mass required to achieve 
loop orbits with a given $p \ge p_{\rm crit}$, at all radii.

For less triaxial haloes, and for a given bulge core density, lower SBH masses 
are required to produce loop orbits below a critical eccentricity.  The minimal
mass of the  bulge needed  to create the required loop orbits  in the outer
region, however, also decreases --- especially since, for a mildly triaxial halo,
the radius at which no loop orbits exist inside the harmonic core  decreases
with decreasing triaxiality.  Thus, in the regime of mild halo triaxiality, the 
ratio of $K$ to (normalized) minimal bulge mass should not depend sensitively on the 
value of $\beta$. We do ignore the fact that triaxiality can be a function of radius, 
and choose a fixed value of $\beta$ for simplicity. 

\subsection{Constraints on morphology and halo properties}

Once orbits of sufficiently large $p$ are present, dissipation will 
reduce radial motion relative to these  orbits, as well as vertical motion
away from the symmetry plane defined by the angular momentum vector. 
Infalling gas will start to populate the newly-formed, round, non-intersecting periodic
loops, leading to the formation of a disk component inside the halo core. At this 
stage the bulge formation stops.
Outside the halo core, gas can accumulate at any stage of the formation process
on closed loop orbits, which always  exist in strongly inhomogeneous
density distributions.  Provided that the halo triaxiality is small, these orbits 
will be nearly circular (for potential axis ratio $\beta$ mildly deviating from unity 
$p \sim \beta$: see, e.g., Rix 1995).

If the halo core radius is exceedingly large, however, it is possible that no
significant disk component will form at all. For if the core is large relative to the total
halo (virial) radius, the  contracting gaseous component will end up in the core, instead 
of spinning up and forming an extended disk. Losing most of its angular momentum to the 
halo, it will eventually end up as part of the bulge component. This would lead to the 
formation of an elliptical rather than a
disk-dominated galaxy. In this context, one expects that haloes with larger core
radii are host to larger spheroidal components. Since  it appears that, in
general, more massive galaxies  are usually of earlier type (e.g., Persic, Salucci \& Stel
1996), one could
deduce that more massive cores have  larger core radii. This would be expected
if these cores followed a Faber-Jackson type relation (for example, if
$M_{\rm H} \sim \sigma_{\rm H}^{4}$, then  $M_{\rm H} \sim R_{\rm H}^{2}$)  
as tentatively suggested  by observations (Burkert 1995; Salucci \& Burkert 2000; 
Dalcanton \& Hogan 2001) and deduced if halo cores formed via the destruction of  
the inner ($\rho \sim r^{-1}$) regions of NFW haloes (Navarro, Frenk \& White 1997). 
In this case, lower mass haloes form when the Universe is denser and are, therefore, more
concentrated. As a consequence, the region where $\rho \sim r^{-1}$ is 
smaller relative to the virial radius for  low mass haloes (El-Zant, Shlosman \&
Hoffman 2001). The existence of a Faber-Jackson type relation for halo cores 
is also required in order to reproduce the $M_\bullet - \sigma_{\rm B}$ relation as 
discussed in the next section.

Within the above  framework, the formation of the SBH is intimately tied to the 
bulge component, whereas the formation of the disk component takes place after 
the processes leading to bulge and SBH formation are essentially complete. The 
bulk of the disk is also expected to form at scales larger than the halo harmonic 
core (but see Section~5). Thus, the deduced correlation involving the bulge and SBH does 
not simply generalize to one involving the disk as well, in 
accordance with observations (e.g., Kormendy \& Gebhardt 2001).

\section{The  $M_\bullet - \sigma_{\rm B}$ relation}
\label{sigma}

The fact that all the orbital properties discussed in this paper depend only on
the relative magnitudes of the masses and spatial scales of the galactic components 
involved has important consequences. For the $M_\bullet-M_{\rm B}$ relationship 
it has the obvious implication that the derived correlation will hold for all halo 
masses. A more powerful prediction transpires in relation to the $M_\bullet-\sigma_{\rm B}$ 
relation. For this to hold, it is necessary that the masses and velocity 
dispersions of halo cores are related in a similar manner.

Suppose that for a given halo there exist unique values for the SBH and bulge masses, 
and for the bulge scalelength. If a Faber-Jackson type relation between the halo core 
mass and 
velocity dispersion exists, a corresponding relation will exist between the bulge 
parameters. It also follows that a similar relation will exist between the SBH mass and 
the bulge (and halo) velocity dispersion.  This can be deduced by simple scaling 
transformations, because the orbital properties we are interested in are all invariant 
with respect to spatial scale and mass transformations. This means that if we 
multiply the masses of the SBH, bulge and halo by some constant $\alpha_{\rm M}$, 
thus effectively changing the mass units, the curves in  Fig.~\ref{homofin} will remain 
invariant. In the same manner, if we multiply all lengthscales by some factor 
$\alpha_{\rm R}$, so that $R_{\rm H} \rightarrow \alpha_{\rm R} R_{\rm H}$ and $R_{\rm B} 
\rightarrow 
\alpha_{\rm R} R_{\rm B}$, thus effectively changing the length unit, all curves in  
Fig.~\ref{homofin} remain the same. This implies that if, for example, $M_{\rm H} 
\propto R_{\rm H}^{2}$, 
then $M_\bullet \propto M_{\rm B} \propto R_{\rm B}^{2} \propto R_{\rm H}^{2}$ are 
equivalent systems 
in the sense decribed above (they have the same axis ratio curves for their loop orbits 
with the same $p$ values as a function of rescaled radius). These systems will all 
follow the $M_{\rm H}-\sigma_{\rm H}$ relation for the halo. In this particular case, 
$\sigma_{\rm H} 
\propto M_{\rm H}^{1/4}$ implies a similar relationship between $\sigma_{\rm B}$ and
$M_{\rm B}$, as well as $M_\bullet$. 

It is of course possible that, for a given halo, the masses of the SBH and bulge, 
as well as the bulge lengthscale, are not unique.  In other words, the subset of 
haloes with a given $M_{\rm H}$ and $R_{\rm H}$ may contain bulges and SBHs with 
a distribution of properties. In Section~3.2 we had assumed that the bulge collapse is 
largely terminated when its density is of the order of the halo core density. This 
in turn fixes $K$, once $p_{\rm crit}$ is determined. In this section we further 
assume that the value of $K$ is not changed as a result of any subsequent bulge growth.
That is, the SBH grows in tandem with the bulge (cf. Section~3.4). This relaxes the 
assumption that the bulge masses are determined by the minimal mass required to create
closed loops with $p \ge p_{\rm crit}$ at all radii, allowing for more 
massive bulges.

One would like to infer to what extent variations in the bulge and 
SBH properies, within a given halo, affect the homology 
relations discussed above. We now show that when one 
accounts for variations of the bulge and the SBH parameters, the departure from the 
aforementioned relation is not dramatic.  This comes about basically because the mass 
and core 
radius of the bulge are correlated, under the assumptions of our model. First, we 
note that the average velocity dispersion of the core of the baryonic component can 
be written as $\sigma_{\rm B}^2 = \alpha G M_{\rm B}/R_{\rm B}$, and the average density 
$\rho_{\rm B} = 2^{-3/2} M_{\rm B}/R_{\rm B}^3$. In this case $\sigma_{\rm B}  \propto 
\rho_{\rm B}^{1/6} M_{\rm B}^{1/3}$. 
Here $\alpha$ depends on the functional form of the density distribution. Its exact 
value is unimportant if all bulges are assumed to have the same functional form for 
their density distributions. Henceforth we set $\alpha=1$.

For a constant bulge-to-halo density ratio
${\hat \rho}_{\rm B}$ a relationship between bulge velocity dispersion and mass 
$M_{\rm B} \propto \sigma_{\rm B}^{3}$ results. In a toy model where the halo density 
does not 
vary at all with radius a constant bulge density determines a unique $K$, given 
$p_{\rm crit}$, and an ``$M-\sigma$'' relation between bulge properties within 
a given halo arises, with index equal  to 3.
Considering more realistic models for the density distribution raises the index 
somewhat. When the core radius of the baryonic component is not very small compared 
to that of the halo, the density of the halo will not be strictly constant in the 
region of interest.  As a result, the slope of the $M-\sigma$ relationship will further 
increase.\footnote{If $\rho_{\rm B}\propto R_{\rm B}^{-\alpha}$, where $\alpha\geq 0$, 
the slope 
of $M_\bullet - \sigma_{\rm B}$ is given by $2 [(3-\alpha)/(2-\alpha)]$ which stays between 
$3-5$ for $\alpha = 0 - 1.5$, and then rises rapidly. The resulting slope of 
$M_\bullet - \sigma_{\rm B}$ is some weighted average of $\rho_{\rm B}$, and greater than 3.} 
This occurs because larger bulges ``see" a smaller mean halo density, and therefore smaller 
bulge densities suffice to produce the same relative contribution to the potential. Since 
$M_{\rm B} \propto \sigma_{\rm B}^3 \rho_{\rm B}^{-1/2}$, an inverse correlation between 
density and $\sigma$ steepens the $M_{\rm B}-\sigma_{\rm B}$ relation. 
(A marginal effect is already seen in Fig.~1 
where the inner maxima produced by bulges with progressively larger ${\hat R}_B$ 
also have progressively larger inner minima of $p$.) 

If the baryonic component giving rise to the gravitational potential is significantly 
flattened there will be an increase in the $M_\bullet - \sigma_{\rm B}$ 
slope to 4. In this case, the surface, rather than the volume density, will determine the 
gravitational field, resulting in a situation where $M_{\rm B} \propto \sigma_{\rm B}^4$.

\begin{figure}[ht!!!!!!!!!!!!!!!!!!!!!!!!!]
\vbox to2.5in{\rule{0pt}{2.5in}}
\includegraphics{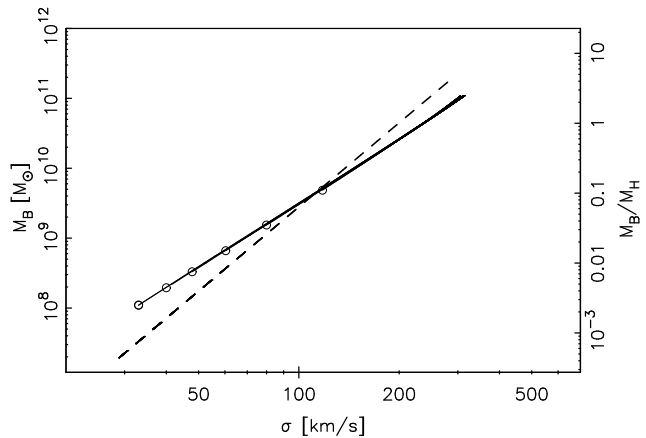}
\caption{Relationship between  bulge mass, $M_{\rm B}$, and velocity dispersion,
$\sigma_{\rm B}$, for systems with ${\hat\rho}_{\rm B} =1$ and for different $p_{\rm crit}$, 
as in Fig.~\ref{compac}. Open circles mark the lower ends of curves corresponding to 
different values of $p_{\rm crit}$, ranging  from 0.3 to 0.8, left to right, and
represent the increase of the minimal bulge mass with $p_{\rm crit}$. The curves for 
different $p_{\rm crit}$ virtually coincide and the circles are placed to denote the minimal
mass, which is a distinguishing feature since it differs with $p_{\rm crit}$. 
The dashed line has a slope of 4. The slope of the solid lines is about 3.2, which 
slightly increases for higher velocity dispersions.
}
\label{comsig}
\end{figure}

The fact that the index of the relation between $M_{\rm B}$ and  $\sigma_{\rm B}$ at constant 
bulge density is close to that of the Faber-Jackson relationship implies that, even 
if there is considerable variation in the bulge mass for a given halo, the departure 
from  a Faber-Jackson relation defined by the halo parameters (as described above) is 
not too large. This  is illustrated in Fig.~\ref{comsig}, where we plot the relationship 
between the average velocity dispersion (simply defined as 
$\sigma^{2}_{\rm B}= GM_{\rm B}/R_{\rm B}$) 
of the baryonic component and its mass  for systems having parameters $K$ and 
$p_{\rm crit}$ corresponding to those in  Fig.~\ref{compac}. The plots  are obtained 
by keeping $K$ and $p_{\rm crit}$ constant and, for a given bulge mass, decreasing 
its characteristic radius until $p \ge p_{\rm crit}$ at all radii inside the halo core. 
To obtain physical parameters we set $V_{\rm H} = 200$ km s$^{-1}$ and $R_{\rm H}=5$ kpc.

The solid line in Fig.~\ref{comsig} is actually a superposition of several lines, 
corresponding to different values of $p_{\rm crit}$ (and $K$), as used in Fig.~2. Its 
average slope is around 3.2 and increases slowly toward higher dispersion velocities. 
These lines, however, have different end-points, determined by the minimal bulge
masses associated with the different $p_{\rm crit}$
(cf. Section~3.3) which are denoted by circles in Fig~\ref{comsig}.\footnote{We note 
here that when the minimal mass is approached, it is necessary to contract 
the bulge by a large amount to maintain $p \ge p_{\rm crit}$ as the mass decreases. 
The result is that small ``hooks" (not shown here) appear 
in place of the circles in Fig.~\ref{comsig}. In principle, this regime may be of 
interest, but is not considered in this paper because it  occurs for a very small 
range of bulge masses.} 
The maximum values of the bulge masses in these curves correspond 
to 2.5 times the halo core mass. It  seems unlikely that bulges would be more massive 
than this. Their characteristic densities or radii would have to be much larger than 
those of the halo,  as would their contribution to the potential. In systems with disks 
such extreme conditions would violate the shapes of observed rotation curves.

\begin{figure}[ht!!!!!!!!!!!!!!!!!!!!!!!!!]
\vbox to2.5in{\rule{0pt}{2.5in}}
\includegraphics{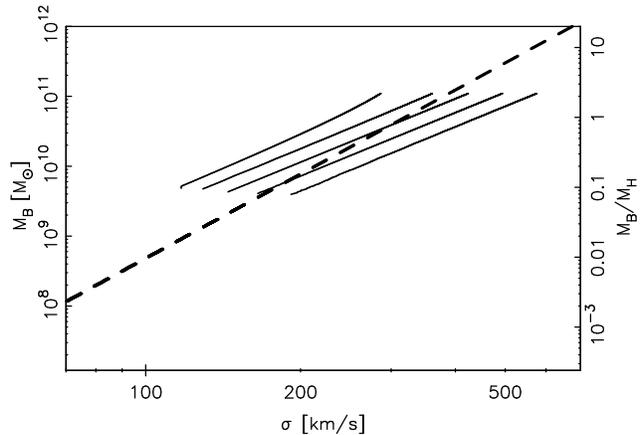}
\caption{Relationship between  bulge mass, $M_{\rm B}$, and velocity dispersion,
$\sigma_{\rm B}$, for systems with $p_{crit}=0.8$ but with different $K$ (and, therefore,
different  densities, as would be clear by following the line with $p_{\rm crit}=0.8$ in 
the scaling regime of Fig.~\ref{compac}). Solid lines, from bottom to the top, are 
$K=10^{-6}$ increasing to $10^{-2}$. The left/right extensions of these lines are 
defined by the limits on ${\hat M}_{\rm B}$ (see text).  The dashed line has a slope of 4. 
The slope of the solid lines is variable, ranging from $\sim 3$ for small $\sigma$ and $K$,
to $\sim 4$ for large values of these parameters.
Note the small dispersion in $\sigma_{\rm B}$ for large variations in $M_{\rm B}$ and $K$.
}
\label{varkmb}
\end{figure}

In reality, for a given halo, there is a range of possible bulge densities and 
associated values of the $K$ parameter. This, in principle, 
affects the normalization of the $M_{\rm B}-\sigma_{\rm B}$ relation.
Nevertheless, the normalization turns out to be  
weakly dependent on ${\hat \rho}_{\rm B}$ ($\sim \sigma_{\rm B}^3 \rho_{\rm B}^{-1/2}$).
Thus, variations in bulge densities within haloes of given mass and size do not 
significantly affect the relationship between bulge mass and velocity dispersion.  
In our model, the central (bulge) density required for the production of loop orbits
with $p \ge p_{\rm crit}$ in the scaling regime is also only weakly dependent on $K$ 
(as $K^{2/5}$, cf. Fig.~\ref{compac}). The above leads to an important corollary, {\it 
that large variation in $K$ will cause only small changes in} $M_{\rm B}-\sigma_{\rm B}$, 
as illustrated in
Fig.~\ref{varkmb}. This relation, therefore, appears to be robust and is not heavily 
affected by changes in bulge and SBH parameters. Note also that, for less dense bulges 
the slope of the lines in Fig.~\ref{varkmb} tends to $\sim 4$, especially in the limit 
of large
values of the velocity dispersion (which is also where most of the observations lie).
This is due to the the effect described above ---  massive bulges with smaller densities 
are more extended, and, therefore probe larger regions of the halo core.

\begin{figure}[ht!!!!!!!!!!!!!!!!!!!!!!!!!]
\vbox to2.5in{\rule{0pt}{2.5in}}
\includegraphics{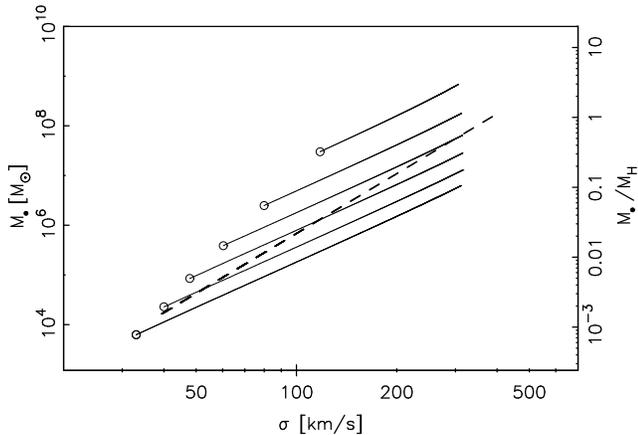}
\caption{Same as in Fig.~\ref{comsig} but for a range of $M_\bullet$,  assuming zero 
scatter in the halo's $M_{\rm H}-\sigma_{\rm H}$ relation. The solid lines are, therefore, 
shifted according to their $K$ values, increasing from bottom to the top. 
These correspond to $K= 5.7 \times 10^{-5}$, $1.2 \times 10^{-4}$, $2.6 \times 10^{4}$,
$5.9 \times 10^{-4}$, $1.6 \times 10^{-3}$ and $K = 6.2 \times 10^{-3}$ (as can be 
inferred from Fig~\ref{comsig}, by considering the intersections of the lines with 
different $p_{\rm crit}$ with the dashed line representing ${\hat \rho}_{\rm B}=1$).
}
\label{varkmbh}
\end{figure}

To obtain an $M_\bullet-\sigma_{\rm B}$ relation, one has to multiply the values of the 
bulge masses in Fig.~\ref{comsig} by the appropriate $K$ factors, arriving at an 
$M_\bullet-\sigma_{\rm B}$ relation {\it within a given halo.} This introduces ``scatter'' 
by shifting the $K={\rm const.}$\ lines, as can be seen from Fig.~\ref{varkmbh}. 
For any fixed value of $\sigma_{\rm B}$, this figure reveals that the 
(vertical) scatter in $M_\bullet$ is similar to that in $K$.

\begin{figure}[ht!!!!!!!!!!!!!!!!!!!!!!!!!]
\vbox to2.5in{\rule{0pt}{2.5in}}
\includegraphics{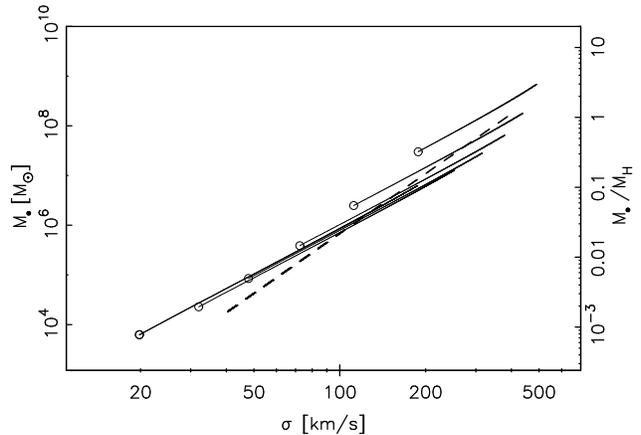}
\caption{Reducing scatter in $M_\bullet-\sigma_{\rm B}$ by $p_{\rm crit}-\sigma_{\rm B}$ 
correlation. A linear correlation between $p_{\rm crit}$ and $\sigma_B$ was imposed in 
Fig.~\ref{varkmbh} (in practice this is done by multiplying each value
of $\sigma_B$ by $p_{\rm crit}/0.5$).  This in effect assumes scatter in the 
halo Faber-Jackson relation, implying that, for any given mass, a range in halo
(and, therefore, bulge) velocity dispersion is plausible and that systems 
embedded in more concentrated haloes are associated with higher values of $p_{\rm crit}$.
The solid lines are, therefore, shifted according to their values of $p_{\rm crit}$, 
significantly reducing the vertical distance between them.
}
\label{shift}
\end{figure}

This leads us to an important point. In the present formulation, the 
$M_{\rm B}-\sigma_{\rm B}$ 
relation is tighter than the $M_\bullet-\sigma_{\rm B}$ one. This appears to 
contradict observations suggesting that $M_\bullet-\sigma_{\rm B}$ is much 
tighter than the Faber-Jackson relation. However, this result has been obtained under 
the assumption that there is {\em no scatter in the halo Faber-Jackson relation.} A 
significant 
scatter in $M_{\rm H}-\sigma_{\rm H}$ would  result in a corresponding 
scatter in $M_{\rm B} -\sigma_{\rm B}$, which follows from the homology 
scaling discussed in the beginning of this section. Moreover, if $p_{\rm crit}$ correlates  
with $\sigma_{\rm B}$ in such a way that gaseous sustems embedded 
in haloes with higher velocity dispersion require larger values of $p_{\rm crit}$ to 
remain stable, the scatter in the $M_\bullet-\sigma_{\rm B}$ relation due to variations 
in $K$ can be reduced. 
For example, in Fig.~\ref{varkmbh}, we have assumed 
a single halo for all the lines with different $p_{\rm crit}$. If, however, 
lines with higher values of $p_{\rm crit}$ are associated with haloes with larger 
$\sigma_{\rm H}$ (and thus $\sigma_{\rm B}$), the constant $p_{\rm crit}$ lines  would 
be shifted in such a way that the vertical distances between them at a given value  
of $\sigma_{\rm B}$ decrease. We illustrate this effect in Fig~\ref{shift}, where a linear 
relationship is assumed between $\sigma_{\rm B}$ and $p_{\rm crit}$. 

We suggest that the relative tightness of $M_\bullet-\sigma_{\rm B}$  
could result from a loose relation between $\sigma_{\rm B}$ and $p_{\rm crit}$, coupled 
with scatter in $M_{\rm H}-\sigma_{\rm H}$ comparable to that of the Faber-Jackson 
relation for bulges. Such $p_{\rm crit}-\sigma_{\rm B}$ relation can result from
a general trend that dissipation is increasing in more concentrated systems.
Testing this will require detailed modeling of the way in which the dissipation rate 
depends on $p_{\rm crit}$, velocity and density, and is outside the scope of this 
paper. Based on the illustrative example of Fig.~\ref{shift}, we only claim here
a plausibility of a loose relationship between $p_{\rm crit}$ and the system velocity 
dispersion. If the postulated correlation 
persists for haloes of different masses, a steepening of the $M_\bullet-\sigma_{\rm B}$ 
relation relative to the halo (and bulge) mass velocity
dispersion relation is expected, implying that the former should have a larger
index.

\section{Discussion and conclusions} 

In the model presented here, gaseous baryonic material settles inside a mildly 
non-axisymmetric halo with a nearly constant density core. Initially, no orbits 
with a definite sense of rotation exist.\footnote{Or only very eccentric  ones if 
one allows for slow figure rotation in the halo.} The first infalling baryonic 
material, therefore, efficiently loses its angular momentum to the core. This initial 
collapse terminates only when the gas becomes self-gravitating and forms stars.  
The increased central density concentration which is produced in this first phase, 
however, destroys the harmonic core, paving the way for the existence of non-intersecting
closed loops with a definite sense of rotation.  

If the loop orbits are too eccentric, the gas will shock and depopulate them. 
These orbits, therefore, cannot represent long-lived attractors of the dissipative
motion. Thus, if the baryonic component does not possess a central cusp initially, the 
absence of sufficiently round supporting closed loops will lead to the formation 
of a central mass 
concentration including an SBH. At larger radii the potential responsible for 
creating sufficiently round loop orbits is that of the extended baryonic component, 
in the form of a bulge. This leads to a linear relationship between the bulge and 
SBH masses. 

It is crucial that both the onset of self-gravity
and the appearance of increasingly circular loop orbits are subject to the same
condition, both depending  on the density ratio of the collapsing gas to that
of the background halo core. This limits the allowable range in densities.
For bulge core densities of order those of the halo core, and for plausible 
values of the critical eccentricity, the value of $K$ lies in the range 
$10^{-4} - 10^{-2}$, which is compatible with present observations.
For  densities up to an order of magnitude larger, the range of values 
is the same, provided that the critical values of the closed loop axis ratios are 
larger. This can be expected if the dissipation rate along loop orbits 
is dependent on density --- a plausible assumption. 

The most important prediction of the model outlined above is that the bulge  and 
SBH  parameters are determined by the halo properties. In particular, 
relationships between SBH and bulge masses and the velocity dispersion
of the bulge necessarily arise, and with the right exponents,  if  the 
haloes should also exhibit a Faber-Jackson type relation between their masses and 
velocity dispersions. Moreover, within a given halo, variations in the bulge and 
SBH properties are not expected to destroy these relations --- because 
imposing the condition of critical eccentricity requires that the bulge and SBH 
masses are related to the bulge velocity dispersion {\em via} a power law, with
index also close to that of the Faber-Jackson relationship. Finally, if the Faber-Jackson
relationship for the halo exhibits significant scatter and if, as again seems plausible, 
the critical
eccentricities  of the loops anti-correlate  with the density of the system,
less scatter should be present in the  relationships between SBH masses and
bulge velocity dispersions than the corresponding relationships between bulges masses and
their own velocity dispersions --- the standard Faber-Jackson relationship.

There is already tentative evidence that halo cores may indeed follow Faber-Jackson type
relationships (Burkert 1995; Dalcanton \& Hogan 2001). In addition, a halo core produced 
by flattening out the inner region (where $\rho \propto r^{-1}$) of the NFW profile 
would produce such a relation (El-Zant, Shlosman \& Hoffman 2001). The model thus 
makes testable predictions concerning  the relationships among the very inner 
regions of galaxies, their extended baryonic components, the dark matter haloes
they are thought to be embedded in and the cosmology that predicts their existence. 

In this framework, larger and more massive halo cores produce, on average, larger 
and more massive bulges. Disks form outside the core, or later, when the central
concentration produced by the baryons destroys the core. If the core is very large, 
most of the baryonic material is consumed in the first phase and no
significant  disk forms. This effect is expected to be prominent in larger mass 
cores, since if these follow the Faber-Jackson relation, more massive
haloes should have  proportionally larger cores. Other predictions include the 
requirement that the  average density of the bulge in the central region
should be close to  that of the halo core. There is also a minimal bulge mass 
associated with a given core, although this varies significantly with the critical 
loop orbit eccentricity assumed. A number of additional consequences for galaxy 
formation and evolution will be discussed elsewhere.

To obtain relationships between SBH masses and bulge properties, we have assumed 
that during the gaseous infall phase the baryonic component did not have a 
central density 
cusp. Most observed bulges and ellipticals, however, do have such cusps. One then 
has to assume that once star formation starts, cold dissipational collapse is 
initiated. This would lead to a central density cusp, as it does in the case of 
cosmological haloes. Memory of the initial state is retained via the total energy, 
which determines the final velocity dispersion.

Merritt \& Ferrarese (2001) have suggested that a ``self-regulating'' mechanism, 
related to the threshold mass necessary for the loss of triaxiality in the system,
may be behind the close correlation between SBH and bulge properties. As these authors 
point out, however, the SBH masses required for strong chaotic behavior leading to 
rapid loss of triaxiality are in fact probably too large --- being of the order of 
a few percent of the mass of the system's baryonic component. This is actually of 
the order of the SBH mass needed to create round loop orbits at {\em all} radii 
inside the halo core. In our model, however, this is not assumed. Instead, an 
additional baryonic component plays the role of creating these orbits in the outer 
region. The collusion between this ``bulge'' component and the SBH in destroying 
the harmonic core and creating a situation whereby stable gaseous motion can exist 
gives rise to the correlations described in this paper.

The correlations obtained here are compatible with the observed ones, and with 
acceptable scatter, despite our lack of knowledge of the values of such parameters as 
$p_{\rm crit}$ and its variation with system properties. Detailed modeling of the gas 
dynamics will be required to further constrain this model. It is also possible that 
our distinction between the dynamical role played by the SBH and that played by the 
bulge core is too restrictive. We have introduced this to be able to obtain quantitative 
results, within the model, solely on the basis of the orbital characterisitics.
In general, the roles of the two components, i.e., the SBH and the bulge, may not be 
too distinct --- formation of the SBH can take place simultaneously with a cuspy bulge. 
For this, the even more ambitious task of a self-consistent treatment, including gas and 
stellar dynamics and star formation, is required. We believe, however, that our results 
are generic and arise from fundamental dynamical phenomena which will manifest themselves 
in any formulation of galaxy formation in mildly triaxial haloes with harmonic cores.

\acknowledgments

We thank our colleagues, too numerous to mention, for stimulating discussions.
This work was supported in part by NASA grants NAG 5-10823, WKU-522762-98-6 and HST 
GO-08123.01-97A (provided by NASA through a grant from the Space Telescope Science 
Institute, which is operated by the Association of Universities for Research in 
Astronomy, Inc., under NASA contract NAS5-26555) to I.S., NSF grant AST-9876887 
to M.C.B., and NSF grant AST-9720771 to J.F.

\appendix

\section{Perturbation analysis}

Let the potential $\Phi$ be a function of the Cartesian coordinates in the halo
symmetry  plane, $\Phi = \Phi (x^{2},y^{2})$ with $\Phi (0,0)=0$. 
We follow de Zeeuw \& Merritt (1983, hereafter dZM) and expand the potential in even 
powers of the coordinates:
\begin{equation}
\Phi =  \frac{1}{2} \kappa_{1}^{2} x^{2} + \frac{1}{2}  \kappa_{2}^{2} y^{2}+
\frac{1}{4} a_{5} x^{4} +  \frac{1}{2} a_{7} x^{2} y^{2} + \frac{1}{4} a_{9} y^{4} +...
\label{pot}
\end{equation}
If no bulge exists, inside the halo's harmonic core only the terms quadratic in the 
coordinates are important. If, due to an additional component, or at the boundaries 
of the core, weak nonlinearity is present the series can be truncated, as above, to 
second order, as higher order terms are unimportant. The coefficients are given by: 
$\kappa_{1}^{2}= 2 \pa_{x^{2}} \Phi$, $\kappa_{2}^{2}= 2 \pa_{y^{2}} \Phi$,
$a_{5} = 2 \pa^{2}_{x^{2}} \Phi$, $a_{7} = 2 \pa^{2}_{x^{2} y^{2}} \Phi$, $a_{9} = 
2 \pa^{2}_{{y}^{2}} \Phi$,  where the derivatives are taken at $x = y = 0$.
In terms of these one defines the auxilliary variables $\mu_{11}=\frac{3}{4} 
\frac{a_5}{\kappa_{1}^{2}}$, $\mu_{12}=\frac{1}{2} \frac{a_7}{\kappa_{1} \kappa_{2}}$,
$\mu_{22}=\frac{3}{4} \frac{a_9}{\kappa_{2}^{2}}$.

The condition for stability of the loop orbits is (see dZM; Table 2A third row)
\begin{equation}
\mu_{12}  (\mu_{11} - \mu_{12} + \mu_{22}) > 0.
\end{equation}
We will be interested in systems that are both mildly nonlinear and mildly 
nonaxisymmetric: thus $\kappa_{1} \sim \kappa_{2}$ and  $a_5 \sim a_9 \sim a_7 < 0$. 
The above condition reduces to
\begin{equation}
a_7 > \frac{3}{2} \Bl\frac{\kappa_{2}}{\kappa_{1}} a_5 + 
    \frac{\kappa_{1}}{\kappa_{2}} a_9\Br,
\end{equation}
which, under the above conditions, is always satisfied. 

It is, therefore, the condition for the existence of loop orbits and not their 
stability that will be of interest to us. In the absence 
of a central mass (SBH) or density cusp these cannot be found arbitrarily close to 
the center, instead there is a {\em bifurcation radius} beyond which 
these exist. This determines the effective core radius of the system.
To second order, the condition for the existence of  loop orbits is
\begin{equation}
\mu_{22} - \frac{1}{2} \mu_{12} \le \frac{\kappa_{2} \delta}{Q} \le \frac{1}{2} 
      \mu_{12} - \mu_{11},
\end{equation}
with $\delta=\kappa_{1}/\kappa_{2} -1$ and (to first order) $Q = H/\kappa_{2}$, where 
$H$ is the Hamiltonian. For $H >0$ (first order potential 
terms dominate in eq.~\ref{pot}) and $\kappa_{1} \la \kappa_{2}$, the above requires
\begin{equation}
H \ge \frac{4 \kappa_{1} \kappa_{2}^{3} (\kappa_{2} - \kappa_{1})}
    {\kappa_2 a_{7} - 3 \kappa_1 a_{9} }.
\label{Hamiltonian}
\end{equation}

In the unperturbed case, the action variables (e.g., Binney \& Tremaine 1987) are 
time-independent and (exact) solutions can be written in terms of these in Cartesian 
coordinates as $x = \sqrt{2 I_{1}/\kappa_{1}} \cos \theta_{1}$, etc. In the mildly 
nonlinear case the solutions of the equations of motion, averaged over a dynamical 
time (denoted below by a ``bar"), approximate the true solution to first order in 
the (relative amplitude of the) perturbation and for a number of dynamical times 
inversely proportional to this (dZM; see also Bogoliubov \& Mitropolsky 1961; 
Arnold 1989). In this case analogous approximate solutions can be given in terms 
of the corresponding action variables. In particular, for loop orbits in mildly 
nonlinear potentials one finds
\begin{equation}
\bar{I}_{1} = \frac{ Q (\frac{1}{2} \mu_{12}- \mu_{22}) + \kappa_{1} - \kappa_{2} } 
     {- \mu_{11} + \mu_{12} - \mu_{22}}
\end{equation}
and
\begin{equation}
\bar{I}_{2} = Q -  \bar{I}_{1}.
\end{equation}
Eliminating $Q$ one gets
\begin{equation}
\bar I_1 = \frac{\bar I_2 +F_2/F_3}{F_1/F_3 -1},
\end{equation}
where $F_1=- \mu_{11} + \mu_{12} - \mu_{22}$, $F_2=\kappa_{1} - \kappa_{2}$  
and $F_3=\mu_{12}- \mu_{22}$.

At bifurcation, to first order, 
\begin{equation}
\bar{I}_{2}= Q = H/\kappa_{2}, 
\label{action}
\end{equation}
where $H$ is given by eq.~(\ref{Hamiltonian}). These orbits are infinitely thin and 
represent oscillations along the $y$-axis, with amplitude 
\begin{equation}
y_{\rm max}=\sqrt{2 \bar I_2/\kappa_{2}}, 
\label{ymax}
\end{equation}
which is the effective core of the bulge-halo system, as mentioned above.  As one 
increases $H$ these become thicker with axis ratio 
$x_{\rm max}/y_{\rm max} = \sqrt{ \frac{\bar I_{1} \kappa_{2} } { \bar I_{2} \kappa_{1} }}$.  
The view we have taken in this paper is that when this ratio
becomes large enough, $\ga p_{\rm crit}$, such loop orbits can support gaseous motion, 
and that stars formed on these orbits can constitute populations
of stellar disks, thus ending the bulge formation stage.  We note here that, beyond the 
bifurcation point, the above relation predicts a rather rapid (as a function of radius) 
transition to round loop orbits. This is confirmed by orbital integration, even though 
if the bulge mass is smaller than a certain minimal mass, in the 
fully nonlinear treatement, the  orbital axis ratio can decrease again.  The second 
minimum in axis ratio curves (e.g., Fig.~1) is thus not reproduced by this 
perturbation analysis. 

\section{Application to  the potential used in this paper}

We now apply the perturbation analysis to the superposition of potentials given by 
eqs.~(\ref{logarithm}) and~(\ref{bulgepot}) with the goal of calculating the bifurcation 
radii of the loops.  This is the effective harmonic core radius of the bulge-halo system.
Since we are only interested in orbit shapes, we set $V_H^2 = G =1$ and use scaled variables, 
as defined in Section~2.  For the halo, we  use 
\begin{equation}
\Phi_H = \frac{1}{2}  \log (1 + x^{2} + q y^{2}),
\end{equation}
where $q$, corresponding to $\beta^{-2}$ in eq.~(\ref{logarithm}), parametrizes the 
nonaxisymmetry. For the bulge, we have
\begin{equation}
\Phi_B = - \frac{\hat M_B}{\sqrt{\hat R_B^{2}+x^{2}+y^{2}}}.
\end{equation}

With the above definitions for the potential we have
\begin{equation}
\kappa_1^{2} =   2 \pa_{x^{2}} \Phi = 1  + \frac{\hat M_B}{\hat R_B^{3}},
\label{kap1}
\end{equation}
\begin{equation}
\kappa_2^{2}=  2 \pa_{y^{2}} \Phi = q +  \frac{\hat M_B}{\hat R_B^{3}}, 
\label{kap2}
\end{equation}
\begin{equation}
a_7 = 2  \pa^{2}_{x^2 y^{2}} \Phi = - q - {3 \over 2}  \frac{\hat M_B}{\hat R_B^5}, 
\end{equation}   
and 
\begin{equation}
a_9 = 2  \pa^{2}_{y^{2}} \Phi = - q^{2} - {3 \over 2}  \frac{\hat M_B}{\hat R_B^5} . 
\end{equation}

From equations~(\ref{Hamiltonian}),~(\ref{action}), and (\ref{ymax}) bifurcation 
happens when the value of the long axis of the loops satisfies
\begin{equation}
y_{\rm max}^{2} =  \frac {8 \kappa_1 \kappa_2 (\kappa_2 - \kappa_1)}{\kappa_2 a_7 - 
    3 \kappa_1 a_9 } .
\label{core}
\end{equation}
Substituting from the expressions above and taking the limits $\hat M_B / \hat R_B^3 
(\approx \hat \rho_B)\gg 1$,  $\hat M_B / \hat R_B^5 \gg 1$, and assuming $q \sim 1$, 
we obtain  
\begin{equation}
y_{\rm max} \approx { 2 (q -1)^{1/2} \over \sqrt{3}} {\hat R_B^{5/2}\over \hat M_B^{1/2}}. 
\label{expo}
\end{equation}
This is the effective harmonic core radius of the bulge-halo system.

\end{document}